\documentclass[submission, Phys]{SciPost}

\hypersetup{
    colorlinks,
    linkcolor={red!50!black},
    citecolor={blue!50!black},
    urlcolor={blue!80!black}
}

\usepackage[bitstream-charter]{mathdesign}
\DeclareSymbolFont{usualmathcal}{OMS}{cmsy}{m}{n}
\DeclareSymbolFontAlphabet{\mathcal}{usualmathcal}

\fancypagestyle{SPstyle}{
\fancyhf{}
\lhead{\colorbox{scipostblue}{\bf \color{white} ~SciPost Physics }}
\rhead{{\bf \color{scipostdeepblue} ~Submission }}

\fancyfoot[C]{\textbf{\thepage}}
}

\usepackage{graphicx}
\usepackage{float}
\usepackage{braket}
\usepackage{amsmath}
\usepackage[normalem]{ulem}
\usepackage[utf8]{inputenc}
\usepackage{amsfonts}
\usepackage{xcolor}
\usepackage{hyperref}

\begin{document}
\pagestyle{SPstyle}

\newcommand{\cmb}[2]{\textcolor{red}{\sout{#1}} \textcolor{blue}{#2}}

\begin{center}{\Large \textbf{
Degeneracy beyond the parity-symmetry protection in the Lipkin-Meshkov-Glick model
}}\end{center}

\begin{center}
Jamil Khalouf-Rivera\textsuperscript{1,2},
Miguel Carvajal\textsuperscript{3,4},
Francisco Pérez-Bernal\textsuperscript{3,4}
\end{center}

\begin{center}
{\bf 1} Departamento de Física, Facultad de Ciencias, Universidad de Córdoba, Campus de Rabanales, Edif. Einstein (C2), Córdoba, 14071, Spain
\\
{\bf 2} Centro de Estudios Avanzados en Física, Matemáticas y Computación, Universidad de Huelva, Huelva 21071, Spain
\\
{\bf 3} Depto. de Ciencias Integradas y Centro de Estudios Avanzados en Física, Matemáticas y Computación, Unidad Asociada GIFMAN CSIC-UHU, Universidad de Huelva, Huelva 21071, SPAIN
\\
{\bf 4} Instituto Carlos I de Física Teórica y Computacional, Universidad de Granada, Granada 18071, SPAIN
\\
* ykhalouf@uco.es
\end{center}


\section*{\color{scipostdeepblue}{Abstract}}
\textbf{\boldmath{%
    Degeneracy patterns in quantum mechanics stem from the system symmetries. In particular, the broken-symmetry phase in the well-known Lipkin-Meshkov-Glick (LMG) model is composed of doubly-degenerate states of different parity. In this work, we show that such doublets can exist even if parity is not conserved. For this purpose, our starting point is an anharmonic LMG Hamiltonian with a second-order ground-state quantum phase transition (GSQPT) and a rich spectrum, with two different excited-state quantum phase transitions. The inclusion in the Hamiltonian of a term inducing a first-order GSQPT breaks the parity symmetry but conserving the exponential degeneracy in the energy doublets. We demonstrate that this phenomenon can be traced back to the existence of a $\mathbb{Z}_2$ symmetry (reflection symmetry) in the system's classical limit phase space that leads to an anti-unitary $\mathbb{Z}_2$ symmetry in the quantum system. 
}}

\vspace{\baselineskip}


\vspace{10pt}
\noindent\rule{\textwidth}{1pt}
\tableofcontents
\noindent\rule{\textwidth}{1pt}
\vspace{10pt}

\section{Introduction}
Ground-state quantum phase transitions (GSQPTs) occur when the ground state of a quantum system undergoes an abrupt change if one or more Hamiltonian control parameters are varied across a critical value~\cite{Carr2010, sachdev_2011}. Such transitions are driven by quantum fluctuations; they occur at zero temperature and pervade different branches of physics~\cite{Carr2010, sachdev_2011}. The GSQPT concept was extended to the realm of excited states with the definition of excited-state quantum phase transitions (ESQPTs)~\cite{Caprio2008, Cejnar2008Pi}, characterized by the occurrence of a nonanaliticity in the system's level density (for a complete list of references in a rapidly-growing field check the recent review Ref.~\cite{esqpt_review}). Such transitions can be reached in two different ways. On the one hand,  one can track  the energy of a given excited state of the system under the variation of one or several Hamiltonian control parameters. On the other hand, ESQPTs can be accessed increasing the system's excitation energy for constant control parameters. In both cases, the structure of the system eigenstates with energies located above and beyond a critical energy value greatly differ. ESQPTs have been studied in different fields, see Ref.~\cite{esqpt_review} and references therein for an updated description of the ESQPTs' state of the art.

The Lipkin-Meshkov-Glick (LMG) model is a widespread and versatile one-dimensional two level  model, originally introduced as a toy model in the field of nuclear physics~\cite{LIPKIN1965188,Meshkov1965199,Glick1965211}, and widely used nowadays to study spin chains with a long-range, all-to-all, interaction. The LMG model allows for the definition of different Hamiltonians to study first-, second-, and third-order GSQPTs \cite{Romera2014,1stqpt,Mazziotti2022}, as well as ESQPTs of different nature~\cite{Franzosi2001,Santos2016,Nader2021,Chinni2021,Gutierrez2021,Gamito2022I,Gamito2022II,Zhang2024}. In particular, the broken-symmetry phase of some LMG model Hamiltonians presents degeneracy patterns that can be understood as parity doublets~\cite{wigner2013group}, a consideration that is confirmed from the system's classical limit~\cite{Castanos2006,CastenPepe2006,Santos2016,1stqpt}. For the sake of clarity, we would like to make emphasis on the fact that for one-dimensional systems as the present one, there is no true degeneracy ---apart from some very particular frameworks \cite{Matamala2010Degeneracy}--- but an exponential approach between adjacent levels~\cite{Un_degeneration}, which is usually known as quasidegeneracy or exponential degeneracy, and it has recently being dubbed as level or spectral kissing in the context of squeezed Kerr resonators~\cite{Chavez2023, Iachello2023, Reynoso2023, Frattini2024,Reynoso2025}. 

A connection with the classical phase-space for the LMG Hamiltonian can be easily established
using the coherent-state (also known as intrinsic-state)
formalism. Such formalism, originally introduced by Gilmore
\cite{Gilmore1979,Ginocchio1980,Bohr_1980,Dieperink1980} gives access to the mean-field limit of
the system \cite{GILMORE1978189}. Such formalism has been used for models having 
  $u(2)$,
$u(3)$,
$u(4)$, and $u(6)$ dynamical algebras~\cite{IBMbook1987,Symmetries1993,ATM1995,LAA_2015}.

One of the authors and a collaborator (FPB and \'Alvarez-Bajo) noticed that the inclusion of an anharmonic term in the two-dimensional vibron model, used for modeling the bending degree of freedom of triatomic molecules \cite{Iachello2003,PBernal2008,Larese2011,Larese2013,KRivera2019,KRivera2020,KRivera2022,Khalouf_anharmonicity}) results in a new ESQPT \cite{PBernal2010}. This result paved the way for the understanding of the isomerization reaction in triatomic molecular species using an algebraic model \cite{KRivera2019,Khalouf_anharmonicity,Rafik2025}. A similar anharmonicity-driven ESQPT was later found in the LMG model~\cite{Gamito2022I}, with a deep influence on the system dynamics \cite{Gamito2022II,Zhang2024} with an associated excited state phase characterized by parity doublets. In this work, we extend these results to a parity-breaking Hamiltonian, with a degenerate phase explained from a  $\mathbb{Z}_2$ symmetry other than parity. Similar results has been shown in the context of superconducting circuits~\cite{1stqpt,KhaloufPRA2026,carrillo2026}.

This work is organized as follows. In Sec.~\ref{theLMGmodel} we introduce the LMG model  and present four different possible Hamiltonian operators, of increasing complexity.  In addition to this, we study the mean-field limit of the system and the discrete symmetries that arise in phase space for the Hamiltonian operators under study. In Sec.~\ref{results} we present the results for a set of quantities used to characterize the ESQPT: the energy gap for adjacent levels, the expectation value of the GSQPT order parameter  and of dynamical algebra generators, and the quantum fidelity susceptibility. This work ends with some concluding remarks in Sec.~\ref{conclusions}. In the appendix~\ref{app:Neffect}, we discuss the influence of the parity of the total number of particles of the system in the ESQPT precursors for the anharmonic parity-symmetric Hamiltonian.

\section{\label{theLMGmodel} The Lipkin-Meshkov-Glick model}
The LMG model, originally introduced in the field of nuclear physics as a toy model to test the quality of different approximations ~\cite{LIPKIN1965188,Meshkov1965199,Glick1965211}, describes a nucleus as a system of fermions that can occupy two levels. Other physical interpretations of the model are  a system of two-level interacting atoms, a high spin system, or an XY Ising model with range-independent all-to-all interactions. In the latter case, assuming a $1/2$-spin chain with $N$ elements, we can express the Hamiltonian in terms of global spin operators $\hat{J}_{\alpha}=\frac{1}{2}\sum_{i=1}^{N}\sigma_{i,\alpha}$, where $\sigma_{i,\alpha}$ are the Pauli matrices for the $i$-th fermion and $\alpha=x$, $y$, and $z$. If we restrict ourselves to the maximum angular momentum representation, $j=N/2$, we reduce drastically the total dimension from $2^N$ to $N+1$~\cite{AFrank_book}.
The system dynamical algebra is $su(2)$, which can be mapped to a bosonic realization  using the Jordan–Schwinger map~\cite{esqpt_review}, that uses bosonic operators $t^{\dagger}$ ($t$), and $s^{\dagger}$ ($s$) to express the four generators of a $u(2)$ Lie algebra as bilinear products of a creation operator with an annihilation one
\begin{align}
    \hat{J}_z&=\frac{1}{2}\left(t^{\dagger}t-s^{\dagger}s\right)=\frac{1}{2}\left(\hat{n}-s^{\dagger}s\right) \nonumber\\
    \hat{J}_+&=t^{\dagger}s \\
    \hat{J}_-&=s^{\dagger}t \nonumber \\
   \hat{N}& =t^{\dagger}t+s^{\dagger}s=\hat{n}+s^{\dagger}s  ~.\nonumber
\end{align}
In this case the
extra generator of the $u(2) = su(2) \oplus u(1)$ algebra, $\hat n_s = s^{\dagger}s$,  does not play any role in
the observed features, but allows for the definition of a total number of bosons
 operator, $\hat{N} = \hat n_s + \hat n_t$. For bosons, the system Hilbert space is restricted to the totally-symmetric $u(2)$ representation, which fixes the total number of particles. For this reason, from now on we omit the operator character of $N$. 


\subsection{Model Hamiltonians}

The LMG model is usually studied using a simple model Hamiltonian
\begin{equation}\label{Hmod}
 \hat{H}_\text{mod}(\xi) = (1-\xi) \hat{n} - \frac{4\xi}{N} \hat{J}_x^2~.
\end{equation}
The control parameter $\xi$
drives the Hamiltonian energy spectrum from a truncated harmonic oscillator, when  $\xi=0$, to a parity-degenerate anharmonic oscillator spectrum, for $\xi=1$. Despite its apparent simplicity, the model Hamiltonian \eqref{Hmod} exhibits a second-order GSQPT and an associated ESQPT that have been  intensively studied \cite{Cejnar_2007_IBM,Santos2016,PRE2017wang,PRA2019wang,LyapPRE2020}. More recently, the effects of the inclusion of a $\hat n^2$ quadratic term have been studied with detail using the Hamiltonian~\cite{Gamito2022I,Gamito2022II},
\begin{equation}\label{aHmod}
 \hat{H}_\text{anh}(\xi,\alpha) = (1-\xi) \hat{n} + \frac{\alpha}{N}\hat{n}\left(\hat{n}+1\right) - \frac{4\xi}{N} \hat{J}_x^2~.
\end{equation}
The included term gives rise to a second ESQPT  for negative values of the anharmonicity parameter, $\alpha<0$.
In previous realizations of the LMG model, besides the fact that $N$ is a constant of  motion, there exists a $\mathbb{Z}_2$ symmetry which splits the
states  into sets with even and odd parity. This symmetry can be described by a rotation of angle $\pi$ around the $z$-axis, $\hat{\Pi}:\left\{J_x,J_y,J_z\right\}$ $\to$ $\left\{-J_x,-J_y,J_z\right\}$ where $\hat \Pi =e^{i\pi(\hat{J}_z+j)}$.

 Now, inspired on~\cite{1stqpt}, we add an additional term to the Hamiltonian, $\hat{V}_{\text{int}}$, that breaks the parity symmetry with an associated parameter, $\delta$ 
\begin{equation}
    \hat{H}_\text{def}(\xi,\alpha,\delta) =     \hat{H}_\text{anh}(\xi,\alpha) + 
  \delta \, \hat{V}_{\text{int}}\label{Hcomplete} ~,
   \end{equation}

\noindent   where
   
\begin{align}
 \hat{V}_{\text{int}}&=-\hat{J}_x +\frac{1}{N}\left(\hat{J}_x\hat{J}_z+\hat{J}_z\hat{J}_x\right)\\
    &=-2\hat{J}_x+\frac{1}{N}\left(\hat{J}_x\hat{n}+\hat{n}\hat{J}_x\right)~.  
\end{align}



Hence, we consider a Hamiltonian with four
interactions and three control parameters. Like other Hamiltonians formulated within the LMG model~\cite{CastenPepe2006,Romera2014}, the present Hamiltonian exhibits both first- and second-order GSQPTs~\cite{1stqpt}. Despite of breaking the parity symmetry for $\delta\neq 0$, Hamiltonian~\eqref{Hcomplete} still keeps a reflection symmetry $\hat{\mathcal{S}}:\left\{J_x,J_y,J_z\right\}\to\left\{J_x,-J_y,J_z\right\}$. To preserve the $su(2)$ closure relation,  
\begin{equation}
    \left[\hat{\mathcal{S}}^\dagger\hat{J}_x\hat{\mathcal{S}},\hat{\mathcal{S}}^\dagger\hat{J}_y\hat{\mathcal{S}}\right]= - \left[\hat{J}_x,\hat{J}_y\right]=\hat{\mathcal{S}}^\dagger i \hat{J}_z \hat{\mathcal{S}}=\hat{\mathcal{S}}^\dagger i\hat{\mathcal{S}} \hat{J}_z ~,
\end{equation}
the considered symmetry must be anti-linear, i.e. does not commute with the imaginary part of complex numbers, $\hat{\mathcal{S}}^\dagger i\hat{\mathcal{S}}=-i$. The anti-linear nature of  $\hat{\mathcal{S}}$ precludes splitting the Hamiltonian  into different blocks due to impossibility of having a matrix representation of anti-unitary operations acting in our Hilbert space~\cite{messiah2020,Kramers1930,Klein1952,Rosch1983}.

\begin{figure}
    \centering
    \includegraphics[width=\textwidth]{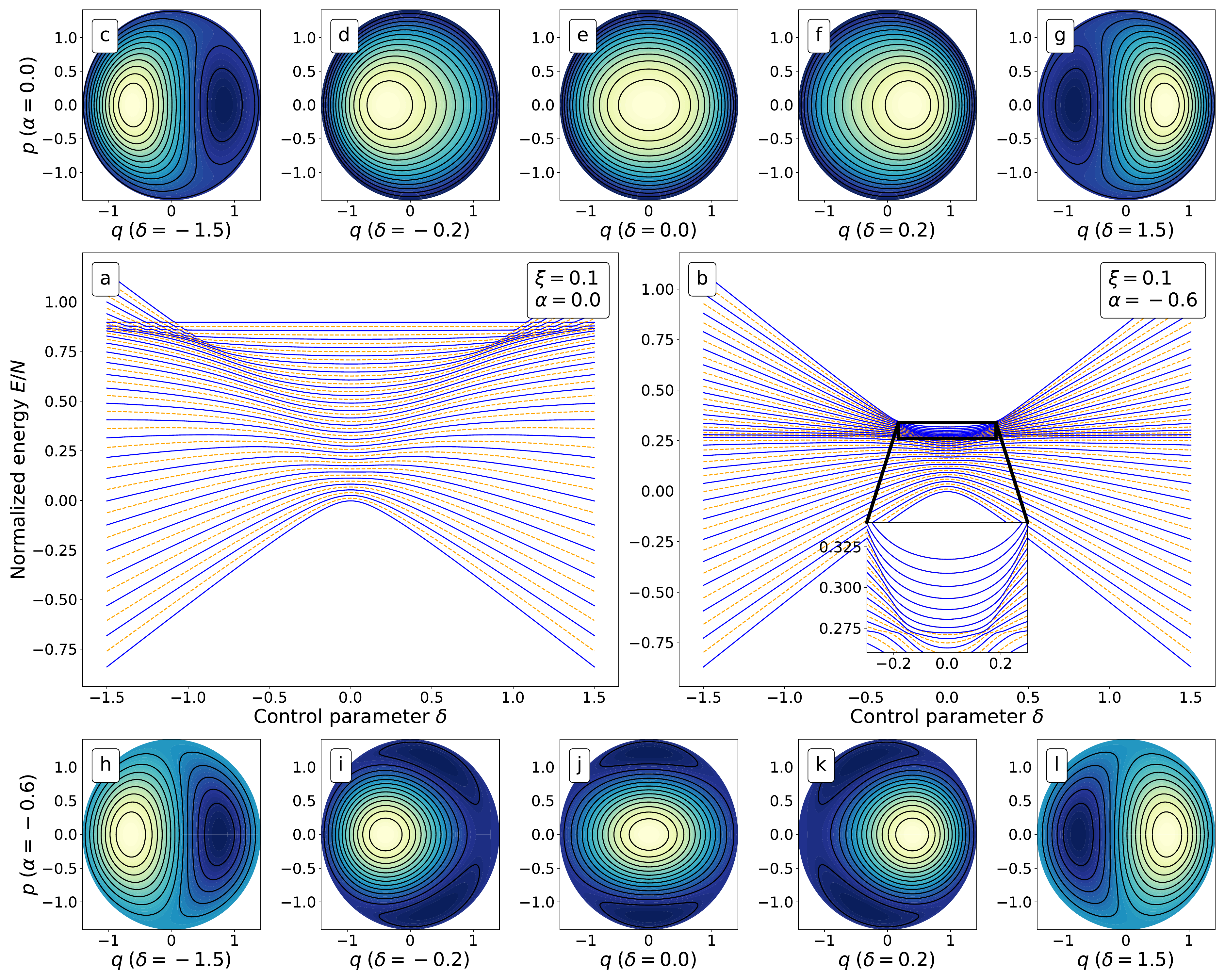}
    \caption{Correlation energy diagram as a function of the $\delta$ control parameter for Hamiltonian \eqref{Hcomplete} with $N=50$, $\xi=0.1$, and  $\alpha=0$ (a) and $-0.6$ (b). The inset of panel (b) is a zoom of the region where the degeneracy appears. For the sake of clarity, different colors and line styles are used to plot adjacent levels. The normalized classical energy surface (see text) for $\alpha=0$ and $\alpha=-0.6$ is depicted in panels (c-g) and (h-l).}
    \label{fig:spectrum0.1}
\end{figure}

\begin{figure}
    \centering
    \includegraphics[width=\textwidth]{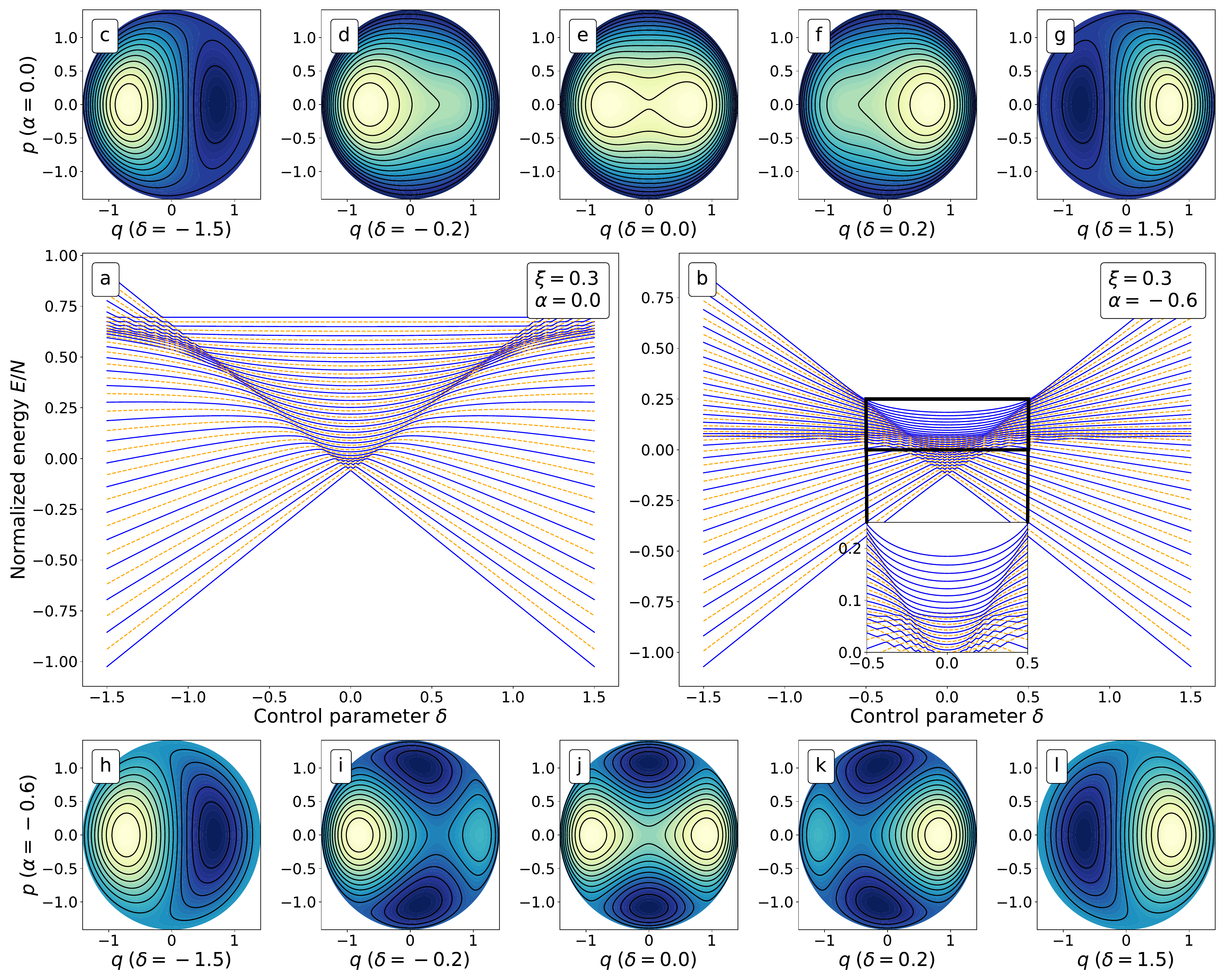}
    \caption{Correlation energy diagram as a function of the $\delta$ control parameter for Hamiltonian \eqref{Hcomplete} with $N=50$, $\xi=0.3$, and  $\alpha=0$ (a) and $-0.6$ (b). The inset of panel (b) is a zoom of the region where the degeneracy appears. For the sake of clarity, different colors and line styles are used to plot adjacent levels. The normalized classical energy surface (see text) for $\alpha=0$ and $\alpha=-0.6$ is depicted in panels (c-g) and (h-l).}
    \label{fig:spectrum0.3}
\end{figure}

The correlation energy diagram of Hamiltonian $\eqref{Hcomplete}$  as a function of the control parameter $\delta$ for $N=50$, $\alpha=0$, and $\xi=0.1$ and $0.3$ is depicted in Figs.~\ref{fig:spectrum0.1}a and \ref{fig:spectrum0.3}a.  For the sake of clarity, alternating line styles and colors are used, though for $\delta\ne 0$ values Hamiltonian cannot be split into blocks. In the spectrum of Fig.~\ref{fig:spectrum0.1}a there is no special feature apart from a high level density  at high energy for  large enough $|\delta|$ values.
However, Fig.~\ref{fig:spectrum0.3}a  is more feature-rich, including level crossing, high level density energies, and a kink for the energy of the ground state at $\delta = 0$. All these features are associated with a first-order GSQPT \cite{1stqpt,Gamito2022I}.

The results for Hamiltonian \eqref{Hcomplete}, with the same parameters than in the previous case and with nonzero $\alpha$  values ($\alpha=-0.6$), are shown in Figs.~\ref{fig:spectrum0.1}b and \ref{fig:spectrum0.3}b. In broad terms, it can be easily noticed that a horizontal line of high density of states appears, as well as a region where states are pair-degenerate (see panel insets) above this line. In the next section we explain the origin of these features through  the analysis of the mean field limit of Hamiltonian  \eqref{Hcomplete}.


\subsection{\label{sec:meanfield} Mean-field limit results}
Hamiltonian \eqref{aHmod} has been already characterized and it presents a  GSQPT at $\xi_c=0.2$,  with a symmetric phase when $\xi<\xi_c$ and a broken-symmetry phase for $\xi>\xi_c$, where the ground state gets doubly degenerate~\cite{Gamito2022I}. In the mean-field limit of the system, the classical Hamiltonian is mapped in the symmetric phase, for $\xi<\xi_c$, to a configuration with a single minimum at the origin, while for $\xi>\xi_c$, in the broken phase, the system has two symmetric minima while the origin is a saddle point. For $\alpha=0$, this saddle point gives rise to an ESQPT at the critical energy $\varepsilon_{c1}$ which, in the large system size limit, defines a separatrix that divides the spectrum into two dynamical phases, a parity-degenerate region for scaled energies $\varepsilon=E/N<\varepsilon_{c1}$ and a non-degenerate one for energies $\varepsilon>\varepsilon_{c1}$. Negative $\alpha$ values give rise to a new ESQPT, with a separatrix at a critical energy $\varepsilon_{c2}$, without altering neither the GSQPT critical point, $\xi_c$, nor the phenomenology of the previous ESQPT. In the large system size limit, the symmetric phase eigenvalues are non-degenerate below $\varepsilon_{c2}$ and parity-degenerate above this energy. In the broken-symmetry phase, energies are
degenerate below and above the two separatrices, and non-degenerate
in-between them. Both values, $\varepsilon_{c1}$ and
$\varepsilon_{c2}$, have been obtained in the mean-field limit of Hamiltonian \eqref{aHmod}
\begin{align}
  \varepsilon_{c1} - \varepsilon_\text{gs}&= \frac{(1-5\xi)^2}{4(4\xi+\alpha)}\;~~~~;~ \xi>\xi_c \\
  \varepsilon_{c2} - \varepsilon_\text{gs}&= \left\{
    \begin{matrix}
      1+\alpha-\xi & ~;~ \xi \leq \xi_c \\
      \frac{(1+2\alpha+3\xi)^2}{4(\alpha+4\xi)} & ~;~ \xi>\xi_c
    \end{matrix}
                  \right.~,
\end{align}
 where $\varepsilon_\text{gs}$ is the ground state energy. A comprehensive analysis
of the LMG model Hamiltonian \eqref{aHmod}  is given in Refs.~\cite{Gamito2022I,Gamito2022II}.

We now study with detail Hamiltonian \eqref{Hcomplete}, including the parity-breaking operator $\hat V_{int}$. The coherent state  for the $u(2)$ algebra  can be defined as
\begin{equation}\label{eq:CohSta}
  \ket{[N],z_s,z_t}=\frac{\left(z_s s^{\dagger}+z_t t^{\dagger}\right)^N}{\sqrt{N!\left(\left|z_s\right|^2+\left|z_t\right|^2\right)^N}} \ket{0}~,
\end{equation}
where $z_s$ and $z_t$ are complex numbers, and  it is normalized for any value of $N$~\cite{Fortunato2010}. We can set $z_s=1$ without loss of information.


The normalized classical limit of Hamiltonian~\eqref{Hcomplete}
can be obtained in the large-size limit $N\to\infty$ by calculating 
its expectation value with the intrinsic state \eqref{eq:CohSta}. In addition, a representation of the classical Hamiltonian in the phase space can be obtained considering 
\begin{align}
    q=&\frac{1}{\sqrt{2}}\frac{\left(z_t+z_t^*\right)}{\sqrt{1+|z_t|^2}} \nonumber \\
    p=&\frac{-i}{\sqrt{2}}\frac{\left(z_t-z_t^*\right)}{\sqrt{1+|z_t|^2}}~,
\end{align}
with $p^2+q^2\leq 2$. Hence, the classical Hamiltonian can be expressed as
\begin{align}\label{eq:Hqp}
    H(q,p)= & \frac{(1-\xi)}{2}\left(q^2+p^2\right)+\frac{\alpha}{4}\left(q^2+p^2\right)^2  \\ + & \xi q^2\left(p^2+q^2-2\right) - \frac{\delta}{2} q\left(2-p^2-q^2\right)^{3/2}~. \nonumber
\end{align}
 The energy surface for the classical Hamiltonian \eqref{eq:Hqp} $\xi = 0.1$ and various $\delta$ values is depicted in Figs.~\ref{fig:spectrum0.1}c-g for $\alpha = 0$ and in Figs.~ \ref{fig:spectrum0.1}h-l for $\alpha = -0.6$. The energy surfaces for the  $\xi = 0.3$ case is depicted in Figs.~\ref{fig:spectrum0.3}c-g for $\alpha = 0$ and in Figs.~ \ref{fig:spectrum0.3}h-l for $\alpha = -0.6$. In all cases the energy is normalized as $\left\{H(q,p)-\min \left[H(q,p)\right]\right\}/\max\left[H(q,p)\right]$ to facilitate the comparison between the different cases.

The stationary points of Hamiltonian \eqref{eq:Hqp} can be obtained from the Hamilton equations,
\begin{align}
    \frac{dq}{dt}&=\frac{\partial H(q,p)}{\partial p}=0 \nonumber \\
    \frac{dp}{dt}&=-\frac{\partial H(q,p)}{\partial q}=0.
\end{align}
 In this case, for $\delta\ne 0$, there is no analytical solution for the stationary points. For the sake of consistency, we denote as $\epsilon_{c1}$ and $\epsilon_{c2}$ the critical energies marking the separatrix at low and high energy values, respectively.

 As already described, for  $\alpha=\delta=0$,  the classical energy surfaces defined by \eqref{eq:Hqp} have either a single minimum at the origin or two symmetric minima and a saddle point at the origin. In Fig.~\ref{fig:spectrum0.1}a, the spectrum of Hamiltonian $\eqref{Hcomplete}$ for $N=50$, $\alpha=0$, and $\xi=0.1$ is depicted as a function of the control parameter $\delta$. For a better understanding of the observed phenomenology, we include the Hamiltonian energy surfaces in panels (c-g) for five different values of the control parameter $\delta$. For $\delta=0$, the system is in the symmetric phase, with a single well centered at the origin. For nonzero  $\delta$ values, parity symmetry is broken and the minimum energy happens at a nonzero $q$ value and $p=0$. For a certain value of $\delta$, a maximum appears at the $q$-axis, with the opposite sign of the minimum, that produces an ESQPT transition with a critical energy equal to the energy value in the system phase space boundary, for $q^2+p^2=2$. 

In Fig.~\ref{fig:spectrum0.1}b we depict the correlation energy diagram as a function of $\delta$ for a system with the same parameter values of the previous case ($N=50$ and $\xi=0.1$) for $\alpha=-0.6$. In this case, the most pronounced feature is a line of high density of states, marking an ESQPT with a critical energy in the large system size limit $\epsilon_{c2}=1+\alpha-\xi=0.3$. The inset in this panel zoomed in a series of doubly-degenerate excited states at energies above $\epsilon_{c2}$ for a range of $\delta$ values around zero. As in the previous case, contour plots for the classical Hamiltonian energy are shown in panels (h-l) for the same selected values of the control parameter $\delta$.

As it is clear when  panels e and j of Fig.~\ref{fig:spectrum0.1} are compared, when $\alpha = -0.6$ there are two maxima with $q=0$. These two maxima are the new maximum value of the system energy, and the energy of the phase-space boundary, $\epsilon_{c2}$, is the critical energy of an ESQPT. When the energy is larger than $\epsilon_{c2}$, the system can lie on any of the two maxima with the same probability, protected by the parity symmetry $\Pi:(q,p)\to(-q,-p)$. When the control parameter $\delta$ is varied,  breaking the parity symmetry, the minimum abandons the origin and the maxima are displaced in the opposite direction. However, the system still conserves the symmetry $\mathcal{S}:(q,p)\to(q,-p)$ and the states above the ESQPT critical energy are doubly-degenerate. At a certain value of $\delta$, the two maxima collapse in one and the degeneracy disappear.

\begin{figure}
    \centering
    \includegraphics[width=\textwidth]{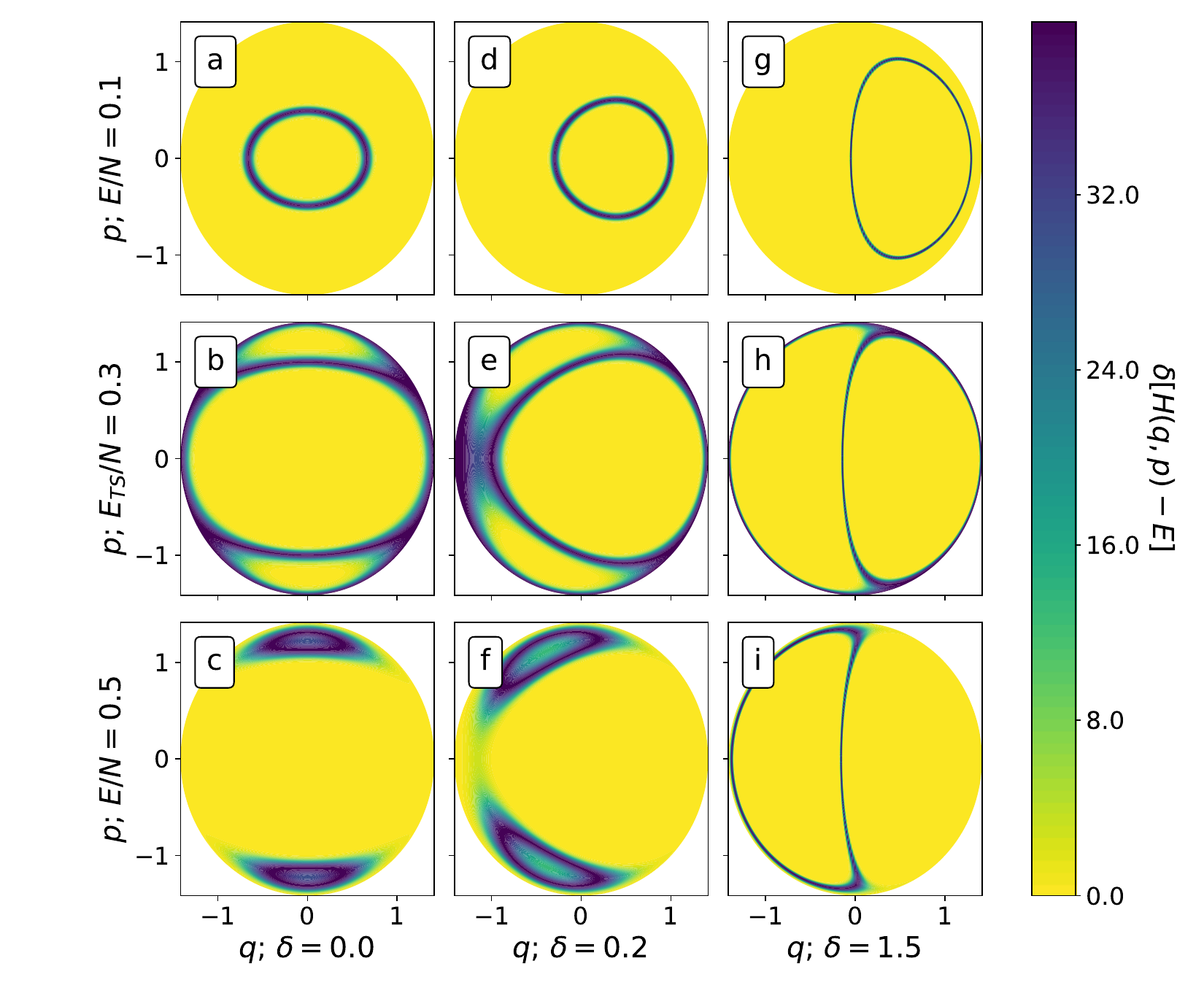}
    \caption{Energy contours of the classical limit of the system \eqref{eq:Hqp} around selected values of the normalized energy  for $\xi=0.1$, $\alpha=-0.6$, and different values of $\delta$. Panels (a-c), (d-f)  and (g-i) correspond to $\delta=0.0$, $\delta=0.2$ and $\delta=1.5$, respectively. Results for normalized energy values below the ESQPT critical energy ($E=0.1$), at the critical energy ($E=\epsilon_{c2} = 0.3$), and above it ($E=0.5$) are depicted in the first, second, and third panel rows, respectively.}
    \label{fig:AS_xi0.1}
\end{figure}

The classical energy functional  \eqref{eq:Hqp} energy contours  depicted in Fig.~\ref{fig:AS_xi0.1} clearly illustrate the different countour shapes at energies  below ($E=0.1$), at  ($E=\epsilon_{c2}=0.3$), and above ($E=0.5$) the ESQPT critical value. The system parameter values are  $\xi=0.1$, $\alpha=-0.6$ and  $\delta=0.0$, $0.2$, and $1.5$. Each panel plots a Dirac delta $\delta(H(q,p;\xi=0.1,\alpha=-0.6,\delta)-E)$ regularized using a narrow bell-shape distribution.
In the first column  ---panels (a-c)--- the system is parity-symmetric, as $\delta=0$. For energies below the ESQPT critical energy (panel a), the system orbits around the minimum, as expected. At the ESQPT critical energy (panel b),  the system explores a large region of its phase space, orbiting around the minimum and the two maxima within the system limit $q^2+p^2=2$. Finally, at energies above the $\epsilon_{c2}$ critical energy  (panel c), the system has an energy above the critical energy $\epsilon_{c2}$ and it orbits around the two symmetric maxima. In the second column ---panels (d-f)--- the control parameter $\delta=0.2$ and the parity symmetry is broken. In this case, the system still presents two maxima, so the phenomenology is very similar to the $\delta= 0$ case, but with orbits that are non-symmetric under $q\to -q$. In the third column ---panels (g-i)---,  for the control parameter $\delta=1.5$, the system orbits around the minimum or the maximum, depending if the energy is lower  ---panel (g)--- or higher  ---panel (i)--- than the critical one. At the critical energy ---panel (h)---, the behavior is similar to panels (b) and (e),  both orbits are connected by the system limit $p^2+q^2=2$. We highlight the fact that the system in all energy regions remains symmetric respect to $p=0$. 

Next we analyze the system for a value of  $\xi\in[\xi_c,1]$, in particular, at $\xi=0.3$. In this case, when $\delta$ and $\alpha$ are zero, the system has parity-degenerate states under the critical energy, which is associated with the saddle point between the two symmetric wells along the $q$-axis, and non-degenerate states above the saddle point at the origin (the energy contour plot for this case is found in Fig.~\ref{fig:spectrum0.3}e). For nonzero $\delta$ values, keeping $\alpha$ equal to zero, parity symmetry is broken and the wells have different depths. A first-order GSQPT occurs for $\delta=0$, where the two wells coexist. This is clearly shown in panels (d) to (f) of Fig.~\ref{fig:spectrum0.3}. If the absolute value of $\delta$ is large enough, one of the wells  disappears and a maximum emerges ---panels  (c) and (g)---.

\begin{figure}
    \centering
    \includegraphics[width=0.9\textwidth]{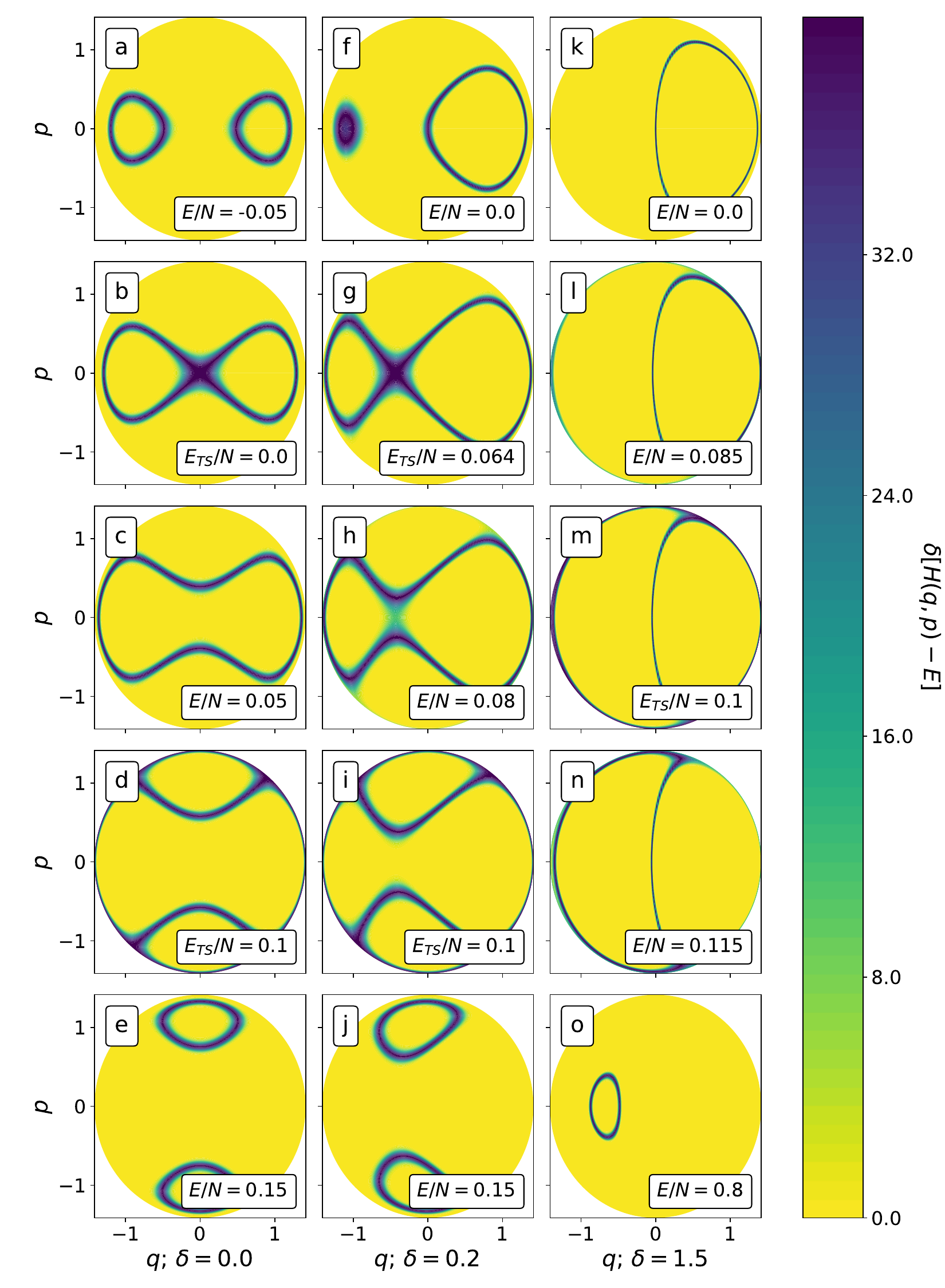}
    \caption{Energy contours of the classical limit of the system \eqref{eq:Hqp} around selected values of the normalized energy  for $\xi=0.3$, $\alpha=-0.6$, and different values of $\delta$. Panels (a-e), (f-j)  and (k-o) correspond to $\delta=0.0$, $\delta=0.2$ and $\delta=1.5$, respectively. Results for normalized energy values at the two ESQPT critical energies are depicted in the second and fourth rows. The other three rows show results for selected normalized energy values above and below the critical energies.}
    \label{fig:AS_xi0.3}
\end{figure}

As shown in  Fig.~\ref{fig:spectrum0.3}, when $\alpha=-0.6$ and $\delta=0$, the system has two minima along $p=0$, separated by a saddle point, and two maxima along $q=0$, separated by the system limit $q^2+p^2=2$ (panel j). For energies lower and higher than both ESQPTs critical energies, states are parity-degenerate; for energies in-between both critical energies, degeneracy is broken. The interplay between the different cases is clear in  panel a-e in Fig.~\ref{fig:AS_xi0.3}. When the scaled energy is $E/N=-0.05$, the classical system remains inside the wells (panel a). At the first ESQPT critical energy, $E_{TS}=0$ (panel b), both wells are connected by the saddle point at the origin. Then, for an energy $E/N=0.05$ (panel c), the system describes a closed orbit in phase space. When the energy is equal to the second ESQPT critical energy, $E/N=0.1$ (panel d), the trajectory in phase space circles the two maxima and both trajectories are connected through the system boundary, for which $p^2+q^2=2$. For values of the energy above the second critical energy, $E/N=0.15$ (panel e), the system describes two disconnected orbits in phase space, around the two symmetric maxima. Summarizing, the existence of two symmetric minima with respect to $q=0$ and two symmetric maxima with respect to $p=0$ explains the existence of energy doublets at energies below the critical energy of the first ESQPT and above the critical energy of the second ESQPT (see inset in panel (b) of Fig.~\ref{fig:spectrum0.3}). In this case, the parity symmetry $\Pi:(q,p)\to(-q,-p)$ can explain the degeneracy. However, we will show that a symmetry $q\to-q$ explains the parity-degenerate states in the wells and a symmetry $p\to-p$ explains the degeneracy found for levels with energy greater than the critical energy of the second ESQPT.

If $\delta$ is non-zero, parity is not conserved anymore, but the system  is still symmetric under $\mathcal{S}$, as shown in panels i and k of Fig.~\ref{fig:spectrum0.3}. This can also be appreciated in panel f to j of Fig.~\ref{fig:AS_xi0.3}).  For low energies, the two minima have different depths (panel f), and at high energies (panel j) the maxima occurs at $q \ne 0$ and  they are still symmetric under $p\to-p$. Therefore the degeneracy below the first ESQPT critical is broken but the energy doublets above the second ESQPT critical energy (panel (b) of Fig.~\ref{fig:spectrum0.3}) are conserved, protected by the anti-unitary $\hat{J}_y$-inversion symmetry in the quantum case. This is clearly evinced in panel j, where the orbits around the maxima are symmetric with respect to $p=0$. For higher values of  $|\delta|$ (e.g., with $|\delta|=1.5$), one minimum disappears and the two maxima collapse in one --- see panels h and l of Fig.~\ref{fig:spectrum0.3}.  The results for this case are included in panels k to o of Fig.~\ref{fig:AS_xi0.3}) and are similar to the ones obtained for a system with $\xi=0.1$, $\alpha=-0.6$ and $|\delta|=1.5$ shown in panels g to i of Fig.~\ref{fig:AS_xi0.1}.

In summary, we have described an excited-state quantum phase in which the deformed LMG model exhibits quasidegenerate doublets protected, in the classical limit, by a symmetry $\mathcal{S}:(q,p)\to(q,-p)$. In the next section, we will demonstrate that such symmetry is linked to $\hat{\mathcal{S}}:\left\{J_x,J_y,J_z\right\}$ $\to$ $\left\{J_x,-J_y,J_z\right\}$ in the quantum system, and that the observed degeneracy is equivalent to the degeneration that occurs in the parity-symmetric case. It is worth clarifying that this degeneracy does not depend on the total number of particles as happens in Kramer's degeneracy \cite{Klein1952} (see App.~\ref{app:Neffect} for a discussion of the parity of the total number of particles in the anharmonic-induced excited-state phase).

\section{\label{results} Degenerate quantum phase under anti-unitary symmetry protection}

\begin{figure}
    \centering
    \includegraphics[width=\textwidth]{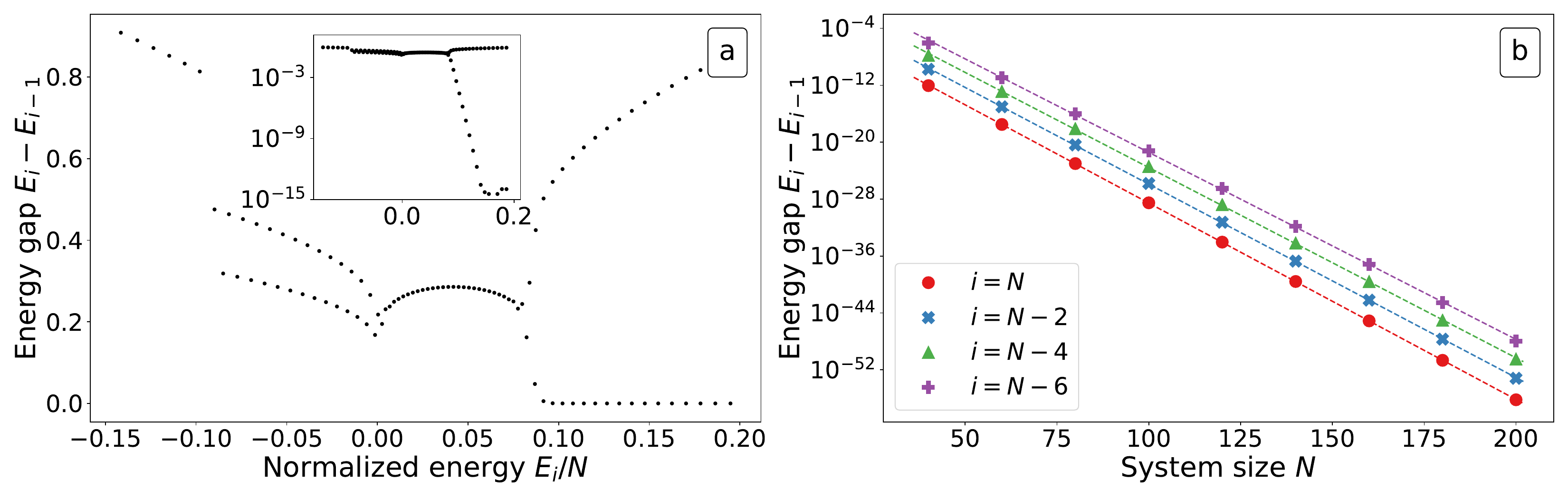}
    \caption{ Panel (a): Adjacent-levels energy gap
    versus normalized energy. In the inset the results are shown using a logarithmic 
    scale in the ordinate axis. Panel (b): Energy gap between the two highest  excited states  versus the system size $N$. The dash lines correspond to the results of a fit assuming the energy gaps are proportional to $e^{-aN}$. All the calculation have been performed for Hamiltonian \eqref{Hcomplete} with $N=100$, $\xi=0.3$, $\alpha=-0.6$, and $\delta=0.05$. Energy gap values are not normalized by the system size.}
    \label{fig:gap_nmean}
\end{figure}

\begin{figure}\centering
  \includegraphics[width=\textwidth]{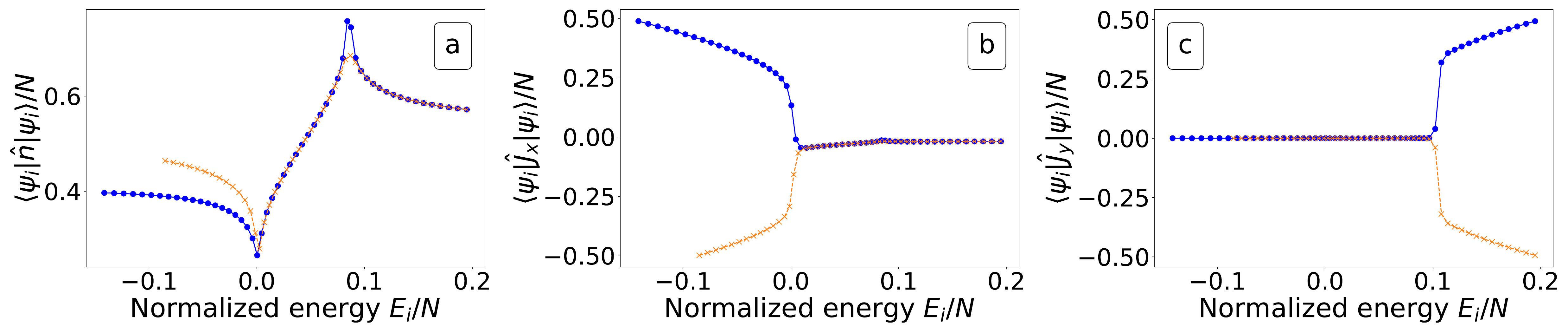}%
  \caption{\label{fig:observables}  Normalized expectation value of the operators $\hat{n}$ ---panel (a)---, $\hat{J}_x$ ---panel (b)---, and $\hat{J}_y$ ---panel(c)--- for a system with $N=100$,
    $\xi=0.3$, $\alpha=-0.6$, and $\delta=0.05$. In panel (c), an interaction $\hat{J}_y$ with a small parameter $\epsilon \simeq ??$ has been added to Hamiltonian \eqref{Hcomplete} (see main text for a detailed explanation).}
\end{figure}

Once the mean-field analysis of Hamiltonian~\eqref{Hcomplete} is completed, we focus on the parameter values where degeneracy appears. In particular, we study a system with control parameter values $\xi=0.3$, $\alpha=-0.6$, and $\delta=0.05$. For such system, the classical Hamiltonian~\eqref{eq:Hqp} has two asymmetric minima, separated by a central saddle point, and two symmetric maxima. All phase space points fulfilling the $q^2+p^2=2$ condition are also stationary points. In order to characterize the system, we calculate the energy
gap between adjacent levels, the expectation value of the operator $\hat{n}$, the expectation value of the projection of the total angular momentum along the $x$- and $y$-axes, and the quantum fidelity susceptibility (QFS). Since the symmetry that protects the degeneracy is antilinear, we cannot separate the Hilbert space into two invariant subspaces , as one would do in parity-symmetric systems~\cite{SolMolPhys1974,Rosch1983,wigner2013group,Geru2018TimeReversal}. However, we will differentiate between two sets of eigenstates, linked to different symmetries.

\subsection{\label{sec:gap-meann}Energy gap}
In the panel (a) of Fig.~\ref{fig:gap_nmean}, we plot
the energy gap of adjacent levels versus the scaled energy for a
system with $N=100$ and the selected control parameter values. When
the normalized energy is below $-0.1$,  states lie
in the  deepest well. For energies in between $-0.1<E/N<0.0$, data split into two branches. For energies
between the saddle point ($E/N=0.0$) and the second ESQPT critical energy 
($E/N\approx 0.1$), the  energy gap between states varies smoothly, switching from positive to negative anharmonicity. Once the second ESQPT energy is crossed the pairs of states at energies $E/N \gtrsim 0.1$ become degenerate. As expected, such degeneracy results in two branches for the energy gap and one of them tends to zero.  

It is well known that a one-dimensional system cannot present pair-degenerate states, except when it has two blocks of non-interacting states (e.g. the isotonic oscillator~\cite{Matamala2010Degeneracy}). 
However, for systems not fulfilling this condition, an exponential approach between adjacent levels can be observed as the system tends to its classical limit. This phenomenon is called quasi-degeneracy and it is responsible for the degeneracy in the broken-symmetry phase of many models, including the LMG model~\cite{Corps2021,Un_degeneration}. 
 For fundamental reasons, this quasi-degeneracy must be protected by some symmetry~\cite{Neumann-Wigner}. To check the nature of the degeneracy patterns observed in the parity-violating Hamiltonian~\eqref{Hcomplete}, we have included an inset in panel (a) of Fig.~\ref{fig:gap_nmean} with the same plot, but using a logarithmic scale for the ordinate axis to verify that the degeneracy is zero up to the numerical precision~\cite{mpmath}. 
In panel (b) of Fig.~\ref{fig:gap_nmean}, we plot using log-lin axes the energy gap between the last two excited states for different system sizes, computed with arbitrary precision~\cite{mpmath}. 
The energy gap values are not normalized by the system size $N$. As expected, the gap decays exponentially as the system size increases, following an exponential law $e^{-aN}$, where the exponent is $a=0.6357\pm0.0005$ for the highest excited state $E_N-E_{N-1}$. 
This is a first hint that this excited-state phase has the same nature than the one observed in the parity-conserving anharmonic LMG model~\cite{Gamito2022I} and the degeneracy is like the one observed in the regular LMG model, despite the fact that parity is not conserved.

\subsection{\label{sec:observables} Projections of the angular momentum}

To understand the behavior of the system in the different dynamical phases, we will study the expectation value of $\hat{n}=\hat{J}_z+j$, $\hat{J}_x$, and $\hat{J}_y$.
The number operator $\hat{n}$ plays the role of an order parameter in the second-order ground-state QPT of the system, as well as it is very useful not only to characterize Hamiltonian~\eqref{aHmod} ESQPTs, but also to classify them: when for eigenstates with energies close to the critical energy of the first ESQPT, the normalized mean value of $\hat{n}$ is minimum and tends to zero with the system size~\cite{Santos2016,Gamito2022I}, while at the critical energy for the second ESQPT, the normalized value of this operator reaches its maximum and tends to one~\cite{Gamito2022I}.

In panel (a) of Fig.~\ref{fig:observables}, we plot the normalized expectation value of $\hat{n}$. In this plot, we distinguish two tendencies, with blue full circles we plot the states belonging to the  absolute minimum, and with orange crosses, the states associated with the local minimum. At low energy values, states are linked to one of the two non-symmetric minima, which is clear examining the two branches below $E/N<0$. At the first ESQPT critical energy, $E/N=0$, this quantity reaches its minimum value. This quantity then grows and reaches its maximum value at the second ESQPT critical energy. This phenomenon is similar to the one observed in Hamiltonian~\eqref{aHmod} (see App.~\ref{app:Neffect} for more details). Once the energy is larger than both critical energies, states are pair-degenerate. In panels (b) and (c), we plot the normalized expectation value of $\hat{J}_x$ and $\hat{J}_y$ using the same color code. In the second case, we have added a perturbation $\epsilon \hat{J}_y$, with $\epsilon = 10^{-6}$ to the Hamiltonian~\eqref{Hcomplete} to break the $\hat{J}_y$-reflection symmetry and make both maxima distinguishable \footnote{This technique is usually used in this kind of situations, for example in Ref.~\cite{DickePedro2017}.}. In panel (b), the mean value of $\hat{J}_x$ exhibits two non-symmetric branches before both transitions and takes a value close to zero for energies larger than the critical energy of the first ESQPT. On the other hand, $\hat{J}_y$ ---panel (c)--- is zero for energies less than the critical energy of the second ESQPT. At this energy, the system starts moving around the two symmetric-maxima with the same probability and the quantum system displays a reflection symmetry $\hat{\mathcal{S}}:\hat{J}_y\to-\hat{J}_y$ that protects the observed exponential-degeneracy patterns despite parity is not conserved. 

\subsection{Quantum Fidelity Susceptibility}
To achieve a better understanding of ESQPT phases in
the deformed model Hamiltonian, we calculate the quantum
fidelity susceptibility (QFS), following the approach presented in
Ref.~\cite{KRivera2022}. Quantum
Fidelity (QF), an important concept in the field of quantum information, is
defined as the module of the overlap between two states
\cite{Nielsen2000,Zanardi2006,Gu2010},
$F=\left|\braket{\phi|\psi}\right|$. If the states depend
on a control parameter $\lambda$, $\ket{\psi(\lambda)}$, the QF can be computed for this state after perturbing the control
parameter,
\begin{equation}
  F(\lambda,\delta\lambda)=\left|\braket{\psi(\lambda)|\psi(\lambda+\delta\lambda)}\right|~.
\end{equation}
Generally, if the perturbation is small enough, this quantity is going
to be close to one for normalized states, except when the system control parameter is close to a critical point, where the nature of the wave function
changes abruptly. The QF depends on the perturbation, $\delta\lambda$, something that can be avoided  calculating the QFS instead, defined as minus the
second derivative respect to the perturbation of the QF,
$\chi(\lambda)=-\frac{\partial^2}{\partial(\delta\lambda)^2}{F(\lambda,\delta\lambda)}$, which is the
leading term in the series expansion of the QF as a function of the
perturbation $\delta\lambda$ \cite{Gu2010,You2007}. The QFS can be computed
for the eigenstates
$\left\{\ket{\phi_j(\lambda)}\right\}_{j=0}^D$ of a system
described by $\hat{H}(\lambda)$ using time-independent
perturbation theory as
\begin{equation}
  \chi_j(\lambda)=\sum\limits_{i\neq j} \frac{\left|\bra{\phi_i(\lambda)}\hat{H}^I(\lambda)\ket{\phi_j(\lambda)}\right|^2}{\left[E_i(\lambda)-E_j(\lambda)\right]^2}~~,
\end{equation}
where $\hat{H}^I(\lambda)$ comes from the expansion of
$\hat{H}(\lambda+\delta\lambda)$ in $\delta\lambda$,
\begin{equation}
    \hat{H}(\lambda+\delta\lambda)\approx
\hat{H}(\lambda)+\hat{H}^I(\lambda)\delta\lambda+\Theta(\delta\lambda^2)~,
\end{equation}
\noindent and $E_j(\lambda)$ is the $j$-th energy. Since the Hamiltonian \eqref{Hcomplete} can have degenerate levels, we need to be careful
applying perturbation theory. 
To overcome this issue, we have diagonalized
$\hat{H}^I(\lambda)$ in the subspace of the degenerate energy $E_j(\lambda)$. This redefines the states as eigenfunctions of both, 
$\hat{H}(\lambda)$ and $\hat{H}^I(\lambda)$. In this way, 
$\bra{\phi_j^{\alpha}(\lambda)}\hat{H}^I(\lambda)\ket{\phi_j^{\beta}(\lambda)}=\delta_{\alpha,\beta}\bra{\phi_j^{\alpha}(\lambda)}\hat{H}^I(\lambda)\ket{\phi_j^{\alpha}(\lambda)}$,
 being $j$ the index of the energy $E_j(\lambda)$, and $\alpha$ and
$\beta$ the indexes of its degenerate states.

Hamiltonian~\eqref{Hcomplete} has three control parameters. Depending on the selected interaction, the QFS will highlight some phenomenon, but not necessary all of them. For this reason, we will use the approach described in Ref.~\cite{KRivera2022}. We are going
to look for a unique control parameter $\lambda$ to compute the
QFS. First of all, we fix the control parameters $\xi$, $\alpha$, and
$\delta$. Then, we divide the Hamiltonian in three parts: (i) a part that commutes with $\hat{n}$, (ii) a second part that commutes with $\hat{J}_x$, and (iii) a term including operators that commute with neither $\hat{n}$ nor $\hat{J}_x$, and multiply them by
the weights $(1-\lambda)$, $(1+\lambda)$ and $(1-\lambda)(1+\lambda)$,
respectively,
\begin{align}
\hat{H}_{\xi,\alpha,\delta}(\lambda)=&(1-\lambda)\hat{H}_{Z}+(1+\lambda)\hat{H}_{X}+(1-\lambda)(1+\lambda)\hat{H}_{\text{mix}}   \nonumber \\
=& (1-\lambda)\left[(1-\xi)\hat{n} + \frac{\alpha}{N}\hat{n}(\hat{n}+1)\right] + (1+\lambda)\left[-\frac{4\xi}{N} \hat{J}_x^2 - 2\delta \hat{J}_x\right] \label{Hlamb} \\ & +(1-\lambda)(1+\lambda)\left[\frac{\delta}{N}\left(\hat{J_x}\hat{n}+\hat{n}\hat{J}_x\right)\right]~~,\nonumber
\end{align}
where
$H^I(\lambda)=-\hat{H}_{Z}+\hat{H}_{X}-2\lambda
\hat{H}_{\text{mix}}$. This method allows us to drive the system between the two limiting cases $\hat H_Z$ and $\hat H_X$,  and recover the original Hamiltonian when the
control parameter $\lambda$ is set to zero.

\begin{figure}
\centering
  \includegraphics[width=0.8\textwidth]{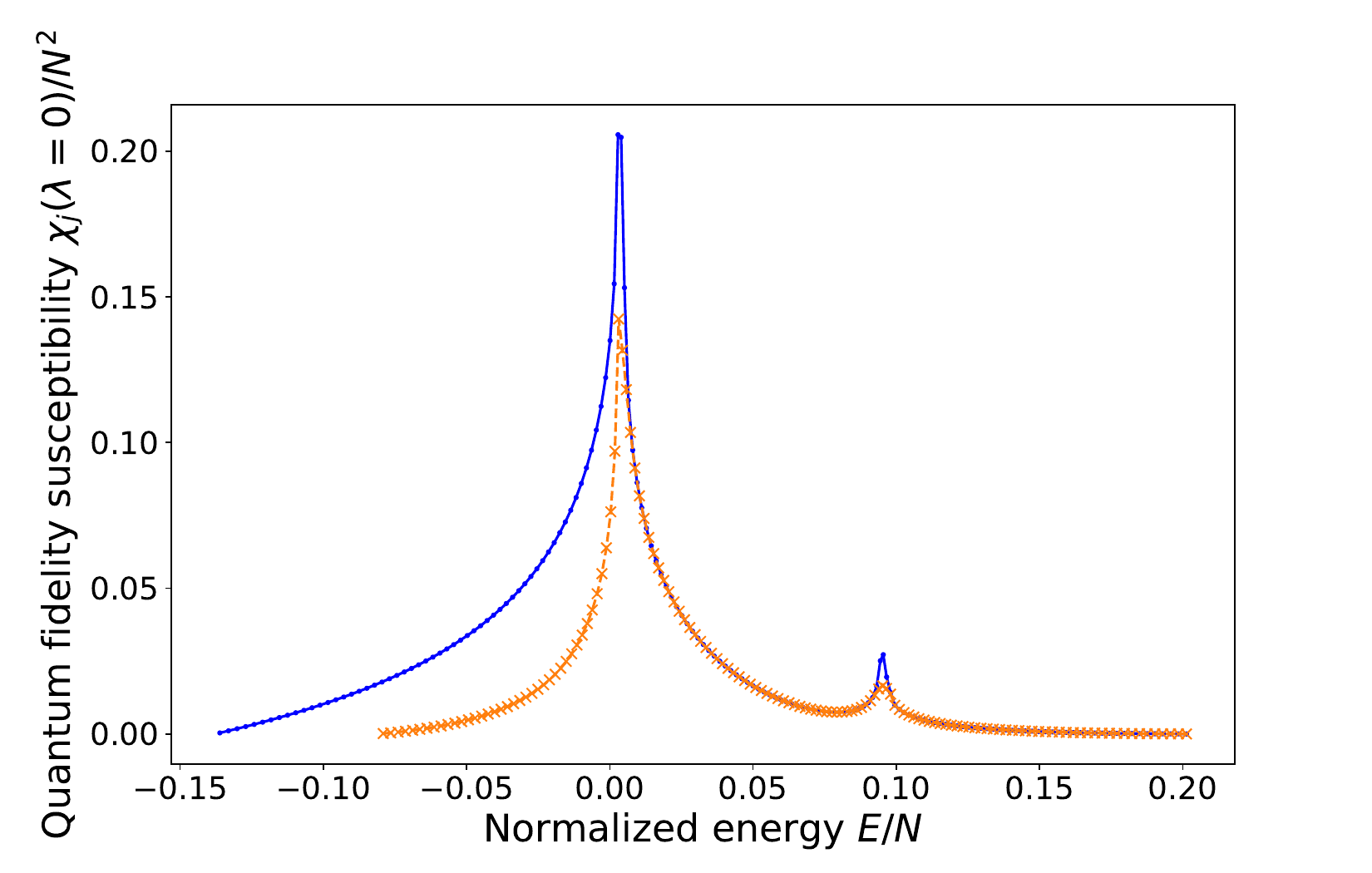}%
  \caption{\label{fig:QFS} Normalized QFS
    versus normalized energy for a system with $N=300$, $\xi=0.3$,
    $\alpha=-0.6$, and $\delta=0.05$.}
\end{figure}

In Fig.~\ref{fig:QFS} we show the normalized QFS (scaled by $N^2$) for
$\lambda=0$ versus the normalized energy for a system with
$N=300$, $\xi=0.3$, $\alpha=-0.6$, and $\delta=0.05$ keeping the same
color code used up to now to distinguish between the two branches. It is clear that the QFS is more sensitive to
the first ESQPT. In a parity-invariant system, the transition due to the central saddle point is more
accessible for states with even parity. Something similar happens with the system-limit ESQPT. In the parity-symmetric problem, even-parity states characterizes better the ESQPT just when the total number of particle $N$ is even (See
App.~\ref{app:Neffect} for an in-depth explanation in the parity-symmetric model
Hamiltonian). In Fig.~\ref{fig:QFS}, we see how the precursors of both ESQPTs are stronger for states associated with the deepest well (blue curve). 
  
\section{\label{conclusions}Conclusions}
In this work, we consider if a parity-breaking LMG model could have non-accidental quasi-degenerate states without conserving the parity symmetry. The answer is yes and, in the case described in this manuscript, such degeneracy is protected by an anti-linear reflection symmetry $\hat{\mathcal{S}}:\left\{J_x,J_y,J_z\right\}\to\left\{J_x,-J_y,J_z\right\}$. This finding can be useful for two main reasons. The first one is because in some quantum computing paradigms, e.g.~quantum annealing, extra interactions may be added to break the parity symmetry and avoid possible degeneracies to work under the adiabatic approximation, or just to use non-degenerate perturbation theory. The second reason is that, considering our results, it is possible to drive a system to a broken-symmetry phase even if parity is not conserved, which could be really useful in those areas where the LMG model helps in the characterization of quantum phase transitions. Although we have presented the results for the LMG model, we can also find degenerate quantum phases in other model, which are protected by an antiunitary symmetry, as is the case of the parity-asymmetric Kerr parametric oscillator~\cite{KhaloufPRA2026}.

Summarizing, we have studied a parity-deformed LMG model which does not conserve the parity symmetry and presents a first-order ground-state QPT and two ESQPTs. We have shown how a pair-degenerate excited-state phase appears. For fundamental reasons \cite{wigner2013group}, we know that there must exist a symmetry to explain the existence of such degeneracy. Using a mean-field analysis, we discovered that the classical limit of our system does not conserve the parity symmetry but conserves the symmetry $p\to-p$, which can be linked to an anti-linear reflection symmetry $\hat{J}_y\to-\hat{J}_y$ in the quantum system. In particular, for some values of the control parameters, two symmetric maxima appear in the mean-field limit, that are equally accessible for the system. These maxima are connected by the classical system-limit.

To be sure about the nature of the degeneracy of this excited-state phase, we show the scaled energy gap between adjacent levels, diagonalizing the system with arbitrary precision. As occurs in the case of parity-symmetric systems, state doublets energy difference decreases exponentially with the system size. We also study different observables and, in all of them, we found two different patterns, as they probe the two existing ESQPTs. In particular, the number operator is a good ESQPT marker as its  expectation value is minimum at the first ESQPT and maximum for the second ESQPT. The expectation value of operators $\hat{J}_x$ and $\hat{J}_y$ helped us to confirm the mean-field analysis results: The system is not invariant under the transformation $\hat{J}_x\to-\hat{J}_x$ but it is invariant under $\hat{J}_y\to-\hat{J}_y$. 

Finally, we studied the QFS and the obtained results are in agreement with the existence of two ESQPTs, confirming the suitability of QFS as an ESQPT probe.

\appendix
\section{\label{app:Neffect}Effect of even and odd number of particles
  $N$ in the regular model Hamiltonian}
In this appendix, we are going to illustrate the effect of the parity of the total
number of particles $N$ over the precursors of the ESQPTs. To this
purpose we are going to use the Hamiltonian \eqref{aHmod}, which has a
good defined parity and is block-diagonal in the number operator  basis.

\begin{figure}\centering

  \includegraphics[width=1.\textwidth]{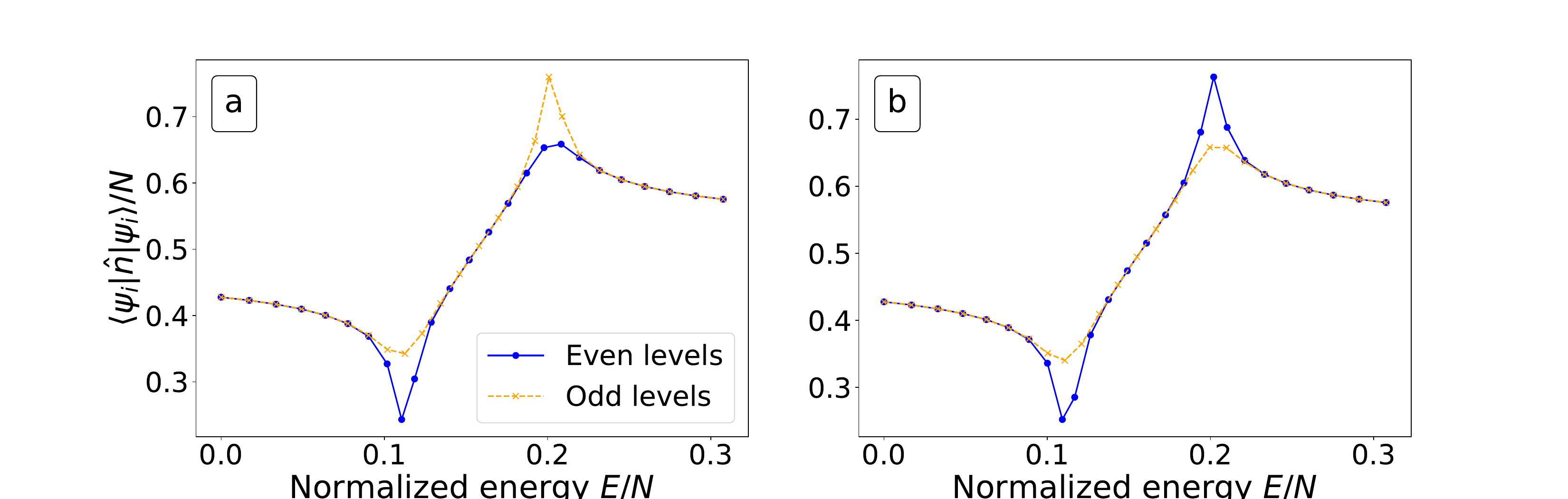}%
  \caption{\label{fig:nmean_49-50} Normalized expectation value of the
    operator $\hat n$ versus the normalized excitation energy for
    Hamiltonian \eqref{aHmod} with different number of particles: $N=49$
    --panel (a)-- and $N=50$ --panel (b)--, and $\xi=0.3$ and
    $\alpha=-0.6$. Blue points correspond to even states and orange
    crosses to odd states.}
\end{figure}

In Fig.~\ref{fig:nmean_49-50}, we plot the normalized expectation value of
operator $\hat n$ versus the normalized excitation energy for systems
with different number of particles, $N=49$ ---odd, panel (a)--- and $50$
---even, panel (b)---, for fixed control parameters $\xi=0.3$ and
$\alpha=-0.6$ of Hamiltonian \eqref{aHmod}. Positive (negative) parity
levels have been plotted using blue points (orange crosses). When the
system goes across the first transition, linked to the central
maximum ($E/N\approx 0.1$), the transition state is well
localized at the first states of the basis ($\ket{n}=\ket{m_z+j}$) basis. Since even and odd
states of the basis are not mixed, the transition state of the
positive-parity block has a stronger precursor of this transition,
since the dominant component is $\ket{0^+}$. On the contrary, the second ESQPT is localized at the last states of the basis,
\begin{equation}
  \begin{matrix}
    \text{Even:} & \ket{0^+} & & \ket{2^+} & &\cdots& \\
    \text{Odd:} & & \ket{1^-}& & \ket{3^-}&\cdots &,
  \end{matrix}   
\end{equation}
which coincides with $\ket{n=N^{\Pi}}$, being the parity $\Pi=+$
($\Pi=-$) when $N$ is even (odd). For even (odd) values of $N$, the
last state of the basis belongs to the positive (negative)
parity states, so the anharmonicity-induced transition precursors will be more
pronounced for positive (negative) states. However, for both numbers of bosons there exists two excited-state phases with degenerate doublets.

\section*{Acknowledgments}
This project received funding through Grant No.~PID2022-136228NB-C21 funded by
MICIU/AEI/10.13039/501100011033 and, as appropriate, by “ERDF A way of making Europe, by ERDF/EU,”
by the European Union, or by the European Union NextGenerationEU/PRTR. This work was also partially supported by the Consejería de Conocimiento, Investigación y Universidad, Junta de Andalucía and European Regional Development Fund, through Grant No. UHU-1262561 (J.K.-R.
and F.P.-B.), and Grant No. PY2000764. This research has also received funding from the European Union’s Horizon 2020 research and innovation program under the Marie Skodowska-Curie Grant Agreement No. 872081. Computing resources supporting this work were
provided by the CEAFMC and Universidad de Huelva High Performance Computer located in the Campus Universitario
“El Carmen” and funded by FEDER/MINECO Project No. UNHU-15CE-2848. 

\bibliography{refs.bib}

\end{document}